# Interface-Driven Growth Mode Control of 2D GaSe on 3D GaAs Substrates with Distinct Crystallographic Orientations

Aida Sheibani[*,1], Mohammad Zamani-Alavijeh[2], Charles Paillard[1,2,3], Kanagaraj Moorthi[2], Fernando Maia de Oliveira[2], Serhii Kryvyi[2], Mourad Benamara[2], Hryhorii Stanchu[2], Calbi Gunder[4], Hugh Churchill[1,2], Yuriy I. Mazur[2], and Gregory Salamo[†1,2]

[1]*Smart Ferroic Materials Center, Physics Department, University of Arkansas, Fayetteville, AR 72701, USA.*

[2]*Institute for Nanoscience and Engineering, University of Arkansas, Fayetteville, AR 72701, USA.*

[3] *Université Paris-Saclay, CentraleSupélec, CNRS, Laboratoire SPMS, 91190, Gif-sur-Yvette, France.*

[4]*Air Force Research Laboratory, 2241 Avionics Circle, Wright-Patterson AFB, OH 45433, USA.*

## ABSTRACT

Previous studies of the growth of two-dimensional (2D) gallium selenide (GaSe) by molecular beam epitaxy (MBE) on a gallium arsenide (GaAs) three-dimensional (3D) substrate have reported significant differences in growth morphology, polytype, and the nature of the interface. The results differ, ranging from GaSe growth in tilted 2D planes to observed spiral structures, thereby calling for a deeper understanding of the impact of the substrate interface on the growth of GaSe films. In this paper, we conduct a comprehensive reexamination of the growth mechanism of GaSe on GaAs substrates with (211)B and (001)B orientations, investigating the nature of the 2D/3D interface and the resulting morphology of the 2D GaSe films. We do this by investigating different methods of preparation of the GaAs substrate surface before the growth of GaSe by MBE, the importance of which has not been considered before. Our results resolve the mechanistic origin of tilted versus non-tilted 2D growth and establish a general interface-driven orientation selection rule linking substrate symmetry and dangling-bond coordination to layered heteroepitaxy. This framework provides a scalable interface-engineering pathway for deterministic control of layered chalcogenide heterostructures and enables wafer-scale integration with established semiconductor device platforms.

Corresponding author:

*asheiban@uark.edu

## 1. Introduction

Of the many possible two-dimensional (2D) materials, gallium selenide (GaSe) is of interest due to its direct band gap producing photoluminescence at ~ 2 eV which can potentially be tuned through thickness control or strain engineering for optoelectronic applications[1, 2, 3]. Moreover, GaSe 2D layers lack inversion symmetry and consequently enable second-order nonlinear optical properties[4-6]. The GaSe layered structure consists of Se-Ga-Ga-Se units bonded via strong in-plane covalent bonds and stacked along the out-of-plane direction through weak van der Waals forces that allow for exfoliation. Unfortunately, the Ga-Ga bonding is weaker than the Ga-Se bond, which results in challenges for exfoliation[7]. For example, although GaSe can be exfoliated from bulk crystals, it often yields films with irregular size and quality[8, 9]. In contrast, direct deposition of GaSe films by molecular beam epitaxy (MBE) has the potential for precise thickness control, uniform coverage, and high quality[10]. However, in this case, the choice of substrate plays a critical role in determining the growth mode, structural quality, and interfacial bonding[11] . Ideally, 2D GaSe films could potentially grow on a substrate without bonding to it[12]. Although 2D materials can potentially act as non-bonding van der Waals substrates, their limited size and availability restrict their use as an MBE substrate. As a result, three-dimensional (3D) substrates, including group-IV semiconductors such as silicon (Si)[13-17] and germanium (Ge)[18], as well as group- III-V semiconductors such as gallium nitride (GaN)[19, 20], and gallium arsenide (GaAs) [21-25], have been investigated as substrates for GaSe growth. For example, Kojima et al.[26] investigated the epitaxial growth of GaSe on GaAs substrates with (001), (111), and (112) orientations and reported GaSe 2D film growth from the surface at tilted angles. Similarly, Sorokin et al.[27] also conducted a study of GaSe growth on GaAs(001) and GaAs(211), but do not confirm tilted growth at the angles observed by Kojima et al.[26] Likewise, Diep et al.[28] investigated GaSe grown on GaAs(001) and observed spiral structures but without tilted growth. While all three investigated the growth of GaSe on GaAs there are significant differences in reported observations. The variability in reported observations calls for a comprehensive investigation of substrate-interface parameters that govern the morphology of a 2D material such as GaSe on a 3D surface, particularly given prior reports that interfacial structure and bonding can strongly influence growth behavior[29]. Interface-driven control of GaSe growth on GaAs substrates also has important implications for device integration. The ability to tailor growth mode and crystallographic orientation provides a route to improved control over interface quality, which is critical for charge transfer, band alignment, and carrier transport in 2D/3D heterojunctions[30, 31]. In particular, GaSe/GaAs structures are promising for optoelectronic applications such as photodetectors and light-emitting devices, where reduced interfacial disorder and controlled carrier transport are essential[32]. Therefore, direct growth of GaSe on technologically mature GaAs substrates provides a viable pathway toward scalable integration of layered semiconductors with existing III–V platforms.

In this paper, we further examine the role of the 3D substrate on the nature of the 2D/3D interface and the morphology of 2D growth. We do this by investigating different methods of preparation of the GaAs substrate surface before the growth of GaSe by MBE, the importance of which has not been considered before. More specifically, through a combination of in-situ reflection high-energy electron diffraction (RHEED), X-ray diffraction (XRD), atomic force microscopy (AFM), transmission electron microscopy (TEM), Raman spectroscopy, and photoluminescence (PL) measurements, we systematically examine the impact of: *(i)* oxide removal, *(ii)* surface preparation with Se, *(iii)* adding a GaAs buffer layer, and *(iv)* growth without oxide removal before the growth of GaSe by MBE on GaAs substrates with (001)B, and (211)B orientations. Our results reveal how substrate preparation and subsequent bonding of the GaSe film with the substrate explain the different outcomes reported in previous studies and add insight on the mechanism for growth of a 2D material on a 3D substrate.

## 2. Experimental Methods

A Riber-32 MBE system equipped with Se and Ga Knudsen effusion cells and an in-situ RHEED system was used to study the growth mechanisms and interface of GaSe films on GaAs substrates. Epi-ready GaAs(B) substrates (Wafer Tech) with (211)B and (001)B orientations were prepared prior to growth. The substrates were solvent-cleaned in acetone and methanol following standard GaAs procedures[33], mounted on a molybdenum holder, and loaded into a load-lock chamber, where they were heated at $200℃$ for 1 hour to remove residual water. The substrates were then transferred to a vacuum degas chamber and thermally cleaned at 350°C for 2 hours. For the current investigation, these prepared epi-ready GaAs(211)B and GaAs(001)B substrates were each divided into four sets before experiments on the growth of GaSe:

***Set 1: Oxide Removed*** - The oxide was thermally removed by introducing the substrate inside the chamber and increasing the temperature into the oxide desorption window ( $580 - 620℃$ ) depending on the substrate orientation, without introducing Se or As flux, and the process was monitored via RHEED intensity. Following oxide removal, rather blurry RHEED reconstruction streaks are observed, indicating a rough GaAs Ga-terminated surface.

***Set 2: Oxide Removed under Se Flux*** - Substrates were ramped from $60℃$ to $350℃$ at a rate of $10℃/min$, after which Se flux was introduced. The substrate temperature was then ramped further into the oxide desorption window ($580 - 620℃$) at which time the RHEED intensity rose and plateaued indicating complete oxide desorption in a Se environment. To ensure complete oxide desorption, the substrates were held at this temperature for 10 min before cooling to the GaSe growth temperature. Prior to initiating GaSe growth on the prepared GaAs substrates, streaky RHEED patterns were obtained with Kikuchi line patterns by maintaining Se flux aiming to create a Se-terminated GaSe surface.

***Set 3: Oxide Removed and GaAs Buffer Added*** - Substrates were transferred to the III-V chamber and ramped in temperature to the oxide desorption window ($580-620$°C), but with an As flux supplied during oxide removal. RHEED monitoring confirmed desorption, after which the temperature was held at the desorption point for 10 min before cooling to a GaAs buffer growth temperature, which was around $580$°C. A GaAs buffer layer was then grown following the procedure of Wangila et al[34] and the sample was transferred to the growth chamber through an ultra-high- vacuum transfer line ($10^{-11}$torr).

***Set 4: Oxide Not Removed*** – The native oxide layer was left on the substrate surface and used for direct GaSe growth on an oxide-covered surface.

Each of these eight GaAs(211)B and GaAs(001)B prepared substrates were then used to grow GaSe under the same growth conditions to examine the role of the surface preparation on the growth morphology and the corresponding nature of the interface.

The beam-equivalent pressures (BEP) were Ga = $0.71 \times 10^{-7}$ Torr and Se =$3.5 \times 10^{-7}$ Torr , giving $\mathrm{Se:Ga} \approx 5:1$. The effusion cell temperatures were $\mathrm{T_{Ga}} = 850$°C and $\mathrm{T_{Se}} = 150$°C. Growth was initiated by opening both shutters of Ga and Se cells at the same time at the growth temperature $\mathrm{T_g} = 400$°C and proceeded for $60$ min for all the samples. For set 3, to suppress As desorption from the surface, we first deposited GaSe for $60\ s$ at $T_g = 375$°C, and then increased to $400$°C for further growth of GaSe for 60 minutes. As a result, the thickness of films grown for 60 min on GaAs(211) is around $104 \pm 14\ nm$ and is around $213 \pm 9\ nm$ when grown on GaAs(001), see Figure S1. All samples reported in this work were mounted on a holding block using indium, and the sample temperature was monitored using a thermocouple in contact with the sample holding block, resulting in good thermal contact between the thermocouple and the sample.

Surface roughness and morphology of the samples were assessed using a Bruker D3100 AFM system (tapping mode, HQ.NSC15/Al BS tips, MikroMasch). XRD measurements were performed using a PANalytical X'Pert MRD system (PANalytical, Almelo, Netherlands) equipped with a 1.6 kW Cu $K\alpha_1$ x-ray source ($\lambda$ = 1.540598 Å) with a vertical line focus, a symmetric 4×Ge(220) monochromator, and a Pixel detector. Raman and photoluminescence (PL) characterization equipment consisted of a Horiba Jobin-Yvon LabRam HR spectrometer, equipped with a Si charge-coupled device detector and a microscope system with a 100x Olympus objective, resulting in an approximately 1 $\mu$m spot size in backscattering geometry. Raman measurements were performed using a 632.8 nm He-Ne laser with an 1800 l/mm grating, while photoluminescence measurements were performed using a 532 nm Nd:YAG laser with a 150 l/mm grating. TEM investigations were performed using an imaging corrected FEI Titan 80-300 microscope operating at 300 kV. The samples were prepared by a classical procedure using mechanical polishing followed by ion milling.

## 3. Results and Discussion

Before discussing the growth of GaSe, it is important to recognize its possible polytypes and polymorphs, since substrate preparation may influence which structure is selected during growth. GaSe is a layered group-III chalcogenide semiconductor that can crystallize in several forms. Most of these are polytypes, which differ only in the stacking sequence of Se-Ga-Ga-Se tetralayers along the crystallographic *c*-axis, whereas the $\gamma'$ phase is a distinct polymorph with a different crystal symmetry, as shown in Figure 1.

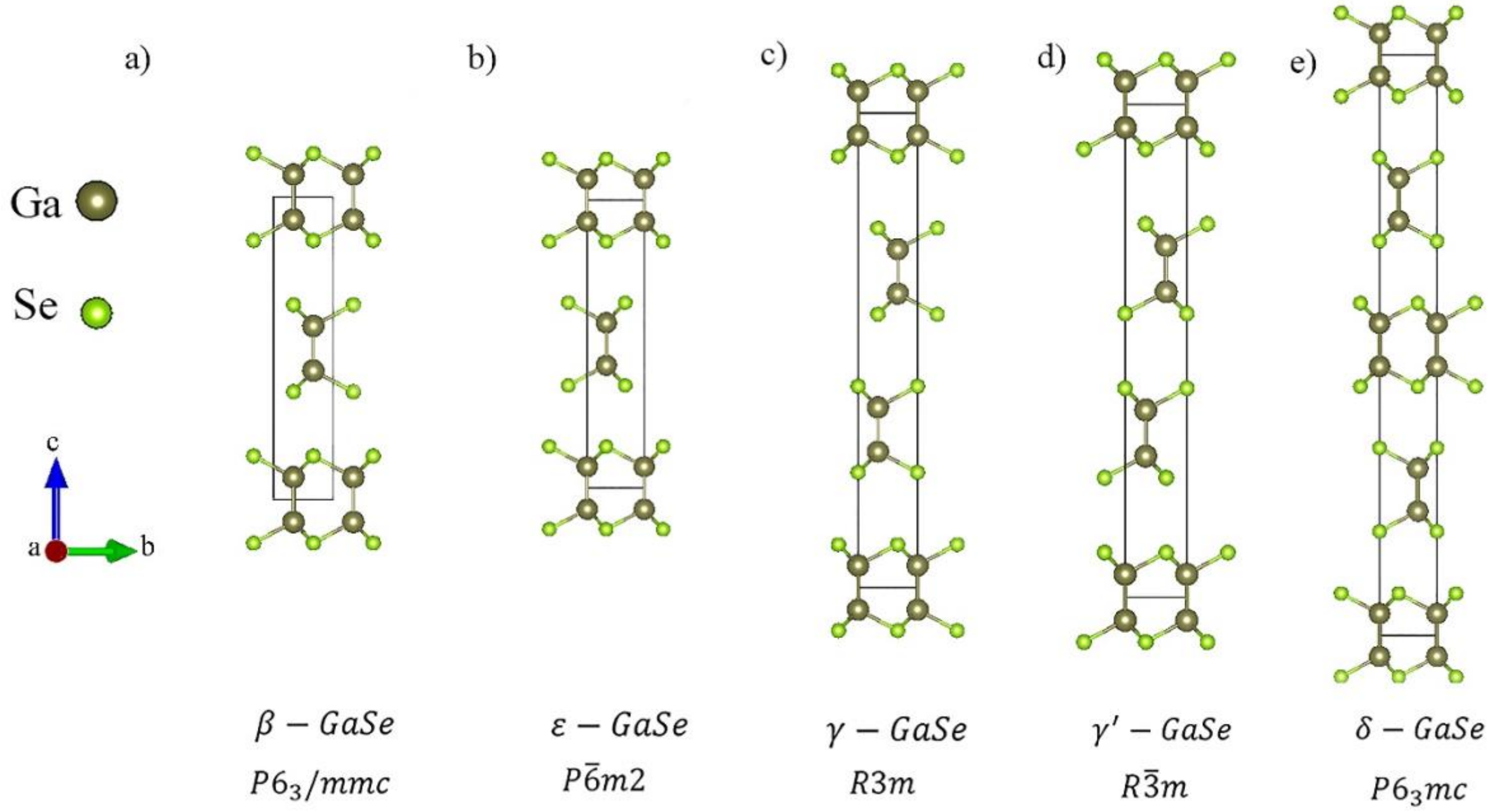

**Figure 1.** Unit cells (black rectangles) of representative GaSe polytypes and polymorphs: (a) $\beta$ −GaSe, (b) $\varepsilon$ −GaSe, (c) $\gamma$ −GaSe, (d) $\gamma'$ −GaSe polymorph, and (e) $\delta$ −GaSe. Gallium atoms are shown in brown and selenium atoms in green. Crystallographic axes are indicated for reference.

The most commonly observed polytypes are the ε-GaSe (hexagonal, space group $P6_3/mmc$) and γ-GaSe (rhombohedral, space group $R3m$) phases, while rarer variants such as γ′-GaSe and β-GaSe have also been reported[13, 19, 22]. Despite sharing nearly identical in-plane lattice parameters, these polytypes differ in their out-of-plane stacking and inversion symmetry. Differences in stacking energetics are only on the order of a few meV per atom, which allows multiple polytypes to coexist at growth interfaces[13, 35]. With these possibilities in mind, we discuss using the four sets of GaAs(211)B and GaAs(001)B prepared substrates to grow GaSe and examine the role and nature of the substrate surface on the GaSe growth morphology and GaSe/GaAs interface.

### 3.1. Growth of GaSe on GaAs(211) with different surface conditions

The GaAs(211) surface is a high-index crystallographic plane that can be described as a corrugated topology composed of alternating terraces and steps (Figure 2a) with the (211) plane intersecting the (111) plane at an angle of 19.46°, that exposes both Ga and As atoms (see Supporting Information, Figure S2). Based on the structural similarity between GaSe and the GaAs(111) plane, we hypothesize that GaSe can nucleate in a bonded configuration along the GaAs(111) direction, leading to 2D film growth at an angle of approximately 20°relative to the GaAs(211) substrate (see Figure 2b). This hypothesis is examined in the following section.

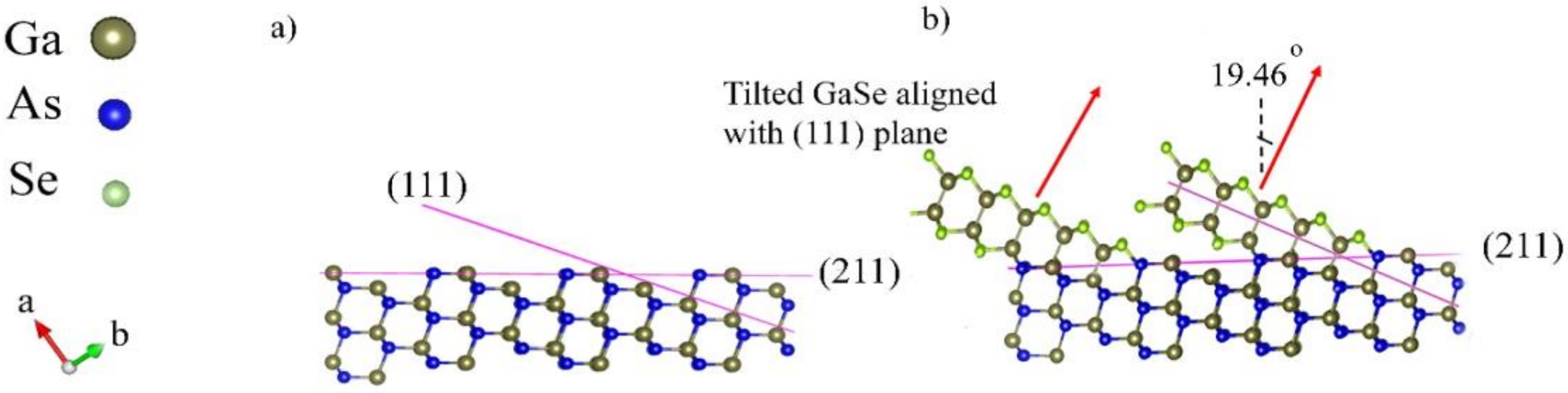

**Figure 2.** Chemical structure of (a) oxide-free and defect-free GaAs(211) substrate; (b) GaSe grown on oxide-free and defect-free GaAs(211) substrate evidencing the angle of 19.46° degrees that GaSe film makes with GaAs(211) substrate. The brown sphere is Ga, the blue sphere is As, and the green sphere is Se. The bond lengths for Ga-As and Ga-Se are nearly the same. The red arrow indicates the direction normal to the (111) surface.

Below, we discuss the growth of GaSe on the four sets of prepared epi-ready GaAs(211) substrates:

***(a) Set 1 - oxide removed*:** Due to the covalent nature of GaAs bonding, oxygen readily forms covalent bonds with either surface As or Ga atoms, depending on the surface termination. On the GaAs(211) surface, where both Ga and As atoms are exposed, both oxides can form although it is likely that only Ga remains after the oxide desorption[36, 37]. Before growth, RHEED from GaAs oxidized surfaces exhibit ring-shaped patterns, characteristic of an amorphous or poorly ordered oxide surface[38]. After removing the oxide layer from the substrate by heating at $580-620$℃ for 10 min, the RHEED pattern (Figure 3a) exhibits a well-defined dotted pattern, characteristic of a rough 3D GaAs surface[39], along with the step feature corresponding to a GaAs(211) surface. At the onset of GaSe growth, immediately after opening the Ga and Se shutters, distinct tilted lines appear in the RHEED pattern (see Figure 3b) and persist throughout the deposition (see Figure 3c). AFM images of the 2D GaSe grown on the oxide-free surface reveal long, tilted structures (see Figure 4b). The RHEED pattern acquired along the $[1\bar{1}\bar{1}]$ azimuth of GaAs(211) shows a tilt angle of $\sim 20^{\circ}$(see Supporting Information, Figure S3). Likewise, the AFM line profile also shows a tilted GaSe 2D film growth with an approximate angle of $\sim 20^{\circ}$ from the GaAs(211) substrate plane along the GaAs (111) direction (see Supporting Information, Figure S4). The outcome, therefore, is suspected to be growth of separated 2D planes at an angle of $\sim 20^{\circ}$ to the substrate as depicted before. To support this hypothesis, we examined XRD. The XRD 2θ-ω scans of GaSe grown on GaAs(211) without a native oxide layer and with the diffraction vector aligned normal to the (211) surface, revealed that there was no GaSe Bragg diffraction (Figure 5a). However, when the sample was tilted in only one specific direction such that the diffraction vector becomes normal to the (111) plane of the (211) substrate, characteristic XRD peaks of GaSe emerged (Figure 5b, orange curve). Additionally, we speculate that the short interruptions observed in the AFM image of the tilted planes are associated with nucleation influenced by the roughness of the initial GaAs surface. The Raman spectra in Figure 5e further support the crystalline growth of GaSe. For example, the Raman spectrum from the tilted 2D films (Figure 5e, orange curve) shows the GaSe characteristic modes of $A_{1g}^{1}\sim 133.2\ cm^{-1}$ , $E_{2g}^{2}\sim 207.0\ cm^{-1}$, $A_{2}^{''}\sim 247.0\ cm^{-1}$, $E_{2g}^{1}\sim 252.0\ cm^{-1}$ and $A_{1g}^{2}\sim 307.0\ cm^{-1}$, consistent with previous reports[20, 28, 40-42]. GaAs modes of TA$\sim 160.0\ cm^{-1}$, TO$\sim 268.0\ cm^{-1}$, and LO$\sim 292.0\ cm^{-1}$ are also observed[43, 44].

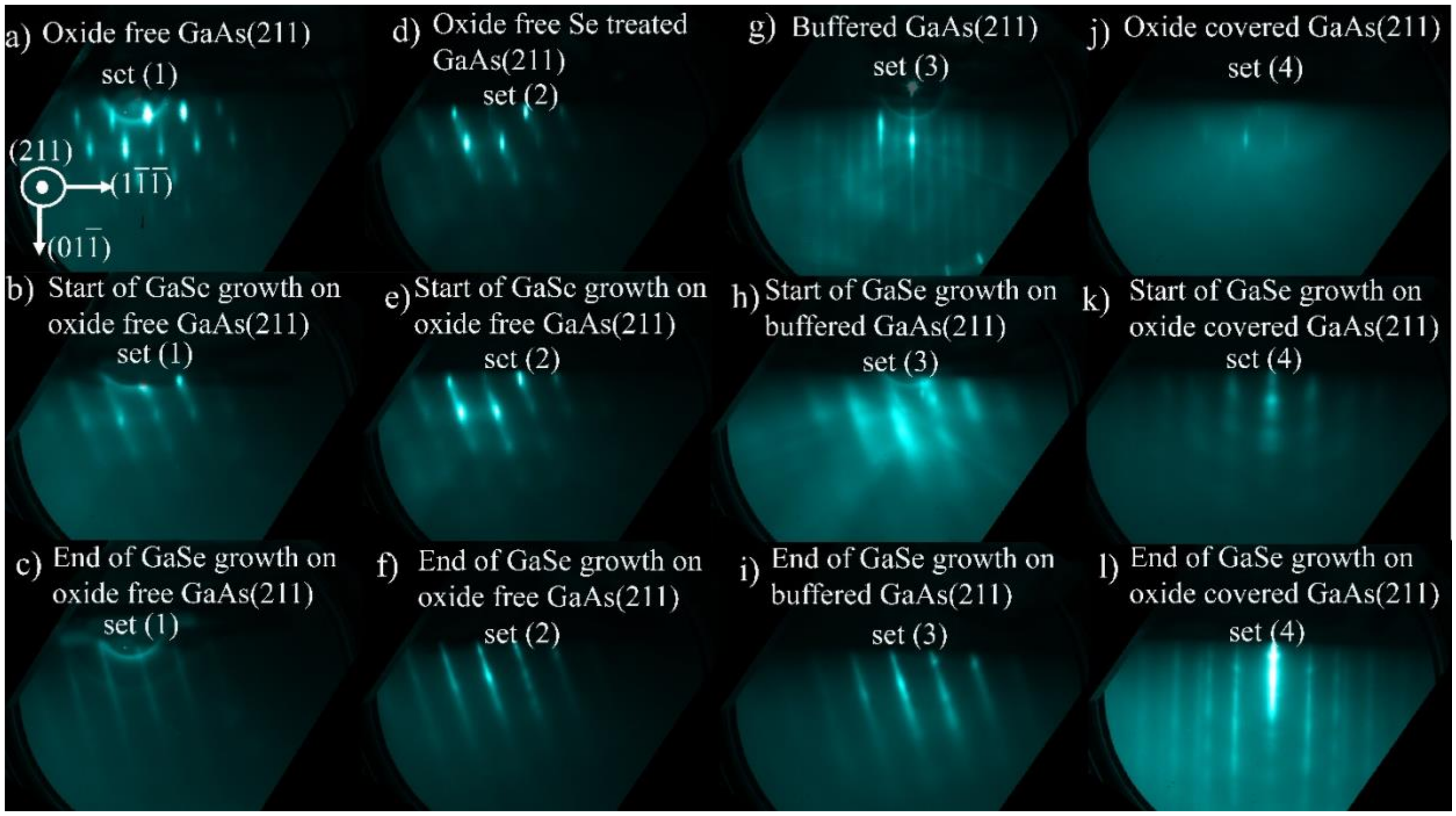

**Figure 3.** RHEED pattern of GaAs(211) substrates and GaSe films under various surface preparation conditions: (a) oxide-free GaAs(211) substrate; (b) beginning of GaSe growth on oxide-free GaAs(211) substrate; (c) end of GaSe growth on oxide-free GaAs(211) substrate; (d) Se treated oxide-free GaAs(211) substrate; (e) beginning of GaSe growth on Se treated oxide-free GaAs(211) substrate; (f) end of GaSe growth on Se treated oxide-free GaAs(211) substrate; (g) buffered GaAs(211) substrate; (h) beginning of GaSe growth on buffered GaAs(211) substrate; (i) end of GaSe growth on buffered GaAs(211) substrate; (j) oxide-covered GaAs(211) substrate, (k) beginning of GaSe growth on oxide-covered GaAs(211) substrate; (l) end of GaSe growth on oxide-covered GaAs(211) substrate.

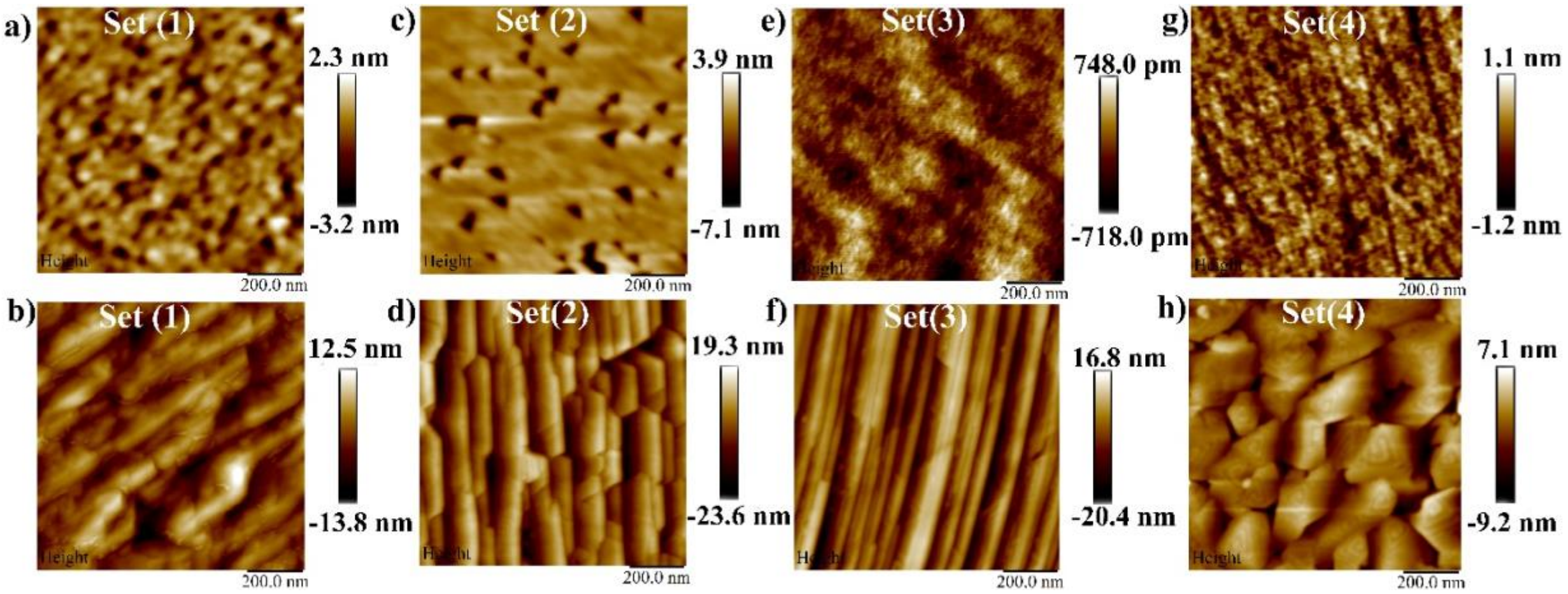

**Figure 4**. Atomic Force Microscopy of surfaces of GaAs(211) substrates and GaSe films under various surface preparation conditions: (a) oxide-free GaAs(211); (b) GaSe film on oxide-free GaAs(211); (c) Se treated oxide-free GaAs(211); (d) GaSe film growth on oxide-free Se treated GaAs(211); (e) buffered GaAs(211); (f) GaSe film growth on buffered GaAs (211); (g) oxide-covered GaAs(211) substrate; (h) GaSe film growth on oxide-covered GaAs(211). All images are 1 μm × 1 μm in size, and the scale bar in each panel represents 200 nm.

***(b) Set 2 - Oxide removed under Se Flux***: Removing the GaAs(211) surface oxide layer in the presence of an excess Se flux enables Se atoms to potentially occupy the resulting surface As vacancies (as indicated by XPS, see Supporting Information, Figure S5 and Figure S6) forming GaSe at the surface of the substrate[45]. Both RHEED and AFM indicate the growth of tilted 2D GaSe planes. However, they appear to be more coherently ordered, marking a clear progression in out of plane crystalline ordering relative to the previous set since the XRD Bragg peaks are sharper and more intense[46] (see Figure 5b, green curve and Figure S7). The key difference from the oxide removal set is that the smoother starting surface (ignoring the triangular distortion) results in nucleation of GaSe layers that are more ordered and aligned, and thus more neatly stacked (see Figure 4d). Our hypothesis is that when GaSe forms at the GaAs substrate surface[45, 47], Se replaces

As or As vacancies at the interface producing a smoother epitaxial starting surface that supports the formation of more ordered 2D GaSe planes. This hypothesis was examined with XPS (Supporting Information, Figure S5 and Figure S6), which indicated that for a GaAs substrate that is exposed to Se before growth of GaSe, there is a decrease in As with a corresponding increase in Se, in agreement with our hypothesis. The Raman spectrum for Set 2 exhibits the same characteristic GaSe phonon modes observed in Set 1 of $A_{1g}^{1}{\sim}133.2\ cm^{-1}, E_{2g}^{2}{\sim}207.0\ cm^{-1}$, $A_{2}^{''}{\sim}247.0\ cm^{-1}$, $E_{2g}^{1}{\sim}252.0\ cm^{-1}$, $A_{1g}^{2}{\sim}307.0\ cm^{-1}$ (see Figure 5e, green curve). The primary distinction is that the $E_{2g}^{1}{\sim}252.0\ cm^{-1}$ shows a noticeably stronger peak relative to the other modes and compared with all other sample sets. The stronger $E_{2g}^{1}{\sim}252.0\ cm^{-1}$intensity observed for set 2 is attributed to the polarization dependence of the Raman response when the laser polarization is perpendicular to the terrace-step structure of the sample. Polarization-dependent Raman measurements indicate that the intensity of this in-plane mode varies with sample rotation relative to the incident laser polarization (see Supporting Information Figure S8).

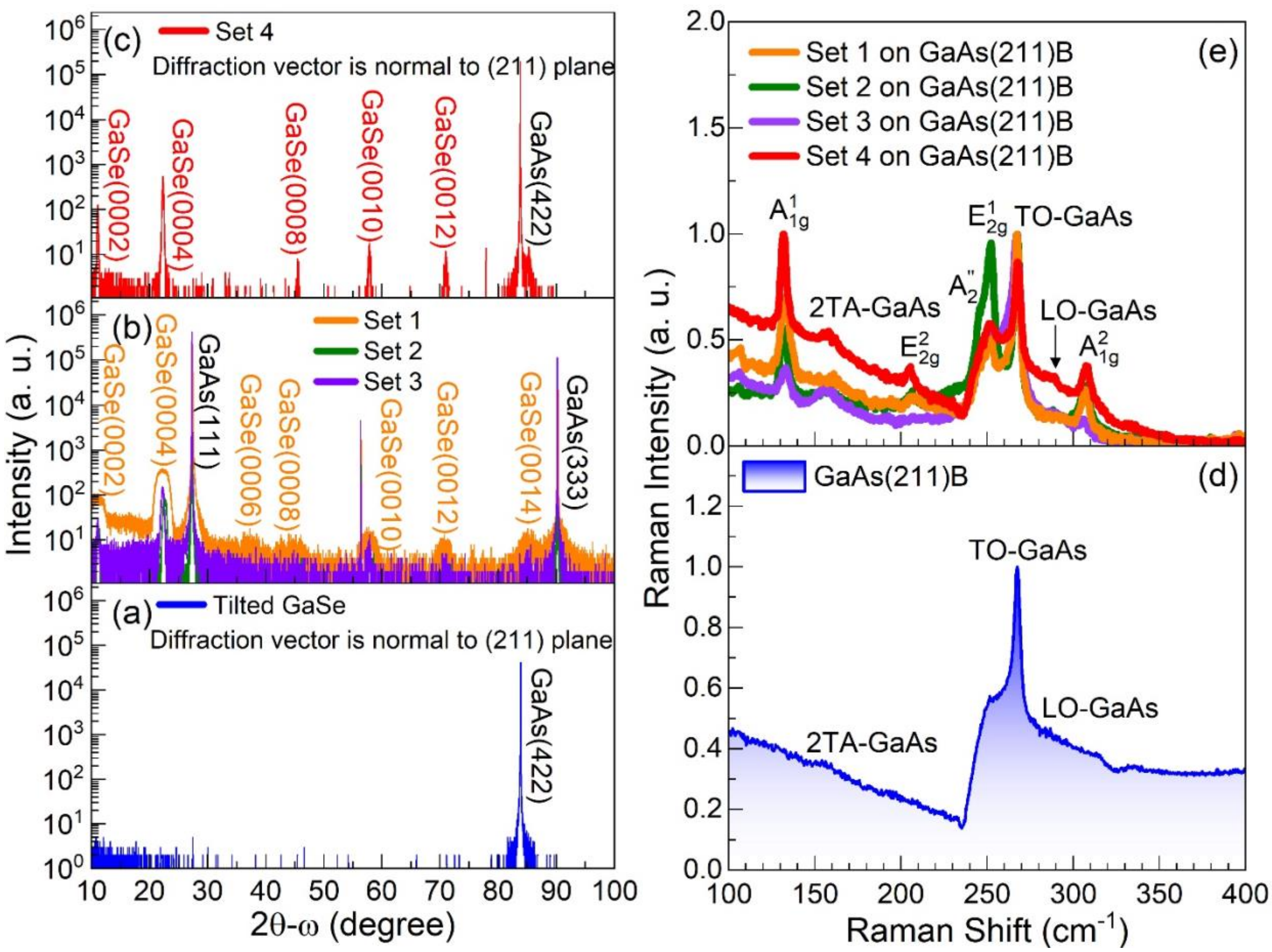

**Figure 5**. Structural and vibrational characterization of GaSe films grown on GaAs(211)B under different surface preparation conditions. (a) XRD pattern of tilted GaSe grown on GaAs(211)B, measured with the diffraction vector normal to the GaAs(211) plane. (b) XRD pattern of tilted GaSe films grown on oxide-free (set-1), selenized oxide-free (set-2), and buffered oxide-free (set-3) GaAs(211)B, measured with the diffraction vector normal to the GaAs(111) plane. (c) XRD pattern of GaSe grown on an oxide-covered GaAs(211)B substrate (set-4), measured with the diffraction vector normal to the GaAs(211) plane. (d) Raman spectrum of the GaAs(211)B substrate. (e) Raman spectra of GaSe films grown on oxide-free (orange curve), selenized oxide-free (green curve), buffered oxide-free (purple curve), and oxide-covered (red curve) GaAs(211)B substrates.

Likewise, direct imaging of the atomic order via TEM in Figure 6 also confirms the crystal quality and tilted growth of the GaSe film at an angle of $\sim 20^{\circ}$ from the substrate interface. Interestingly, certain regions exhibit a ε-GaSe structure although defects are present (see the red rectangle in the inset of Figure 6a). Overall, we observe an outcome similar to set-1 except for the more ordered array of 2D GaSe planes, consistent with our hypothesis of the impact of surface smoothness on coherent 2D growth.

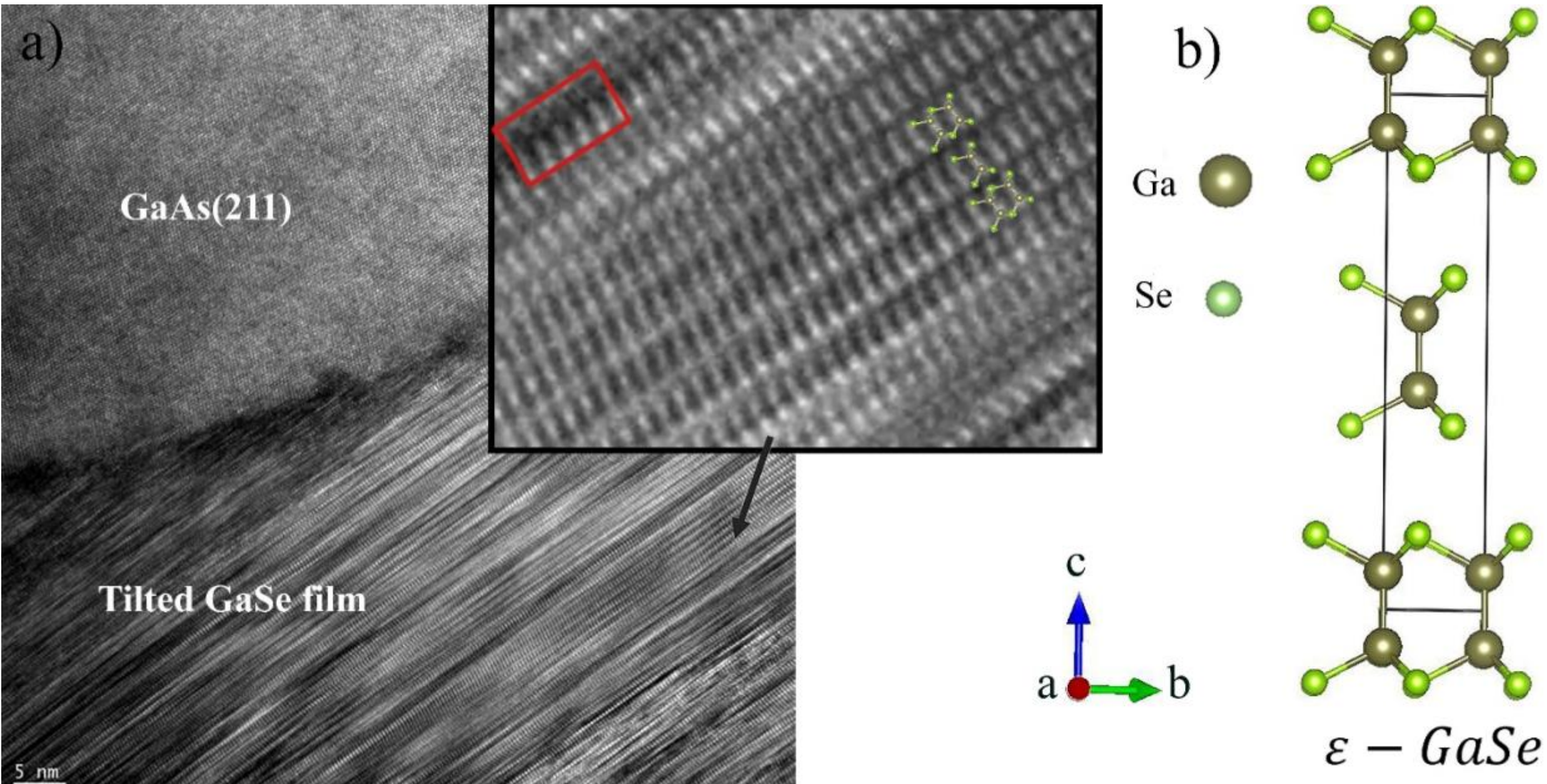

**Figure 6.** (a) Cross-sectional TEM image showing tilted two-dimensional GaSe growth on oxide-free GaAs(211). The inset shows a magnified region where the atomic stacking sequence corresponding to the ε-GaSe polytype is observed. The black rectangle highlights the region used for local polytype identification, while the red rectangle indicates the presence of a stacking fault. (b) Crystal structure and unit cell of ε-GaSe for reference.

***(c) Set 3 - Oxide removed and GaAs buffer added***: Depositing a GaAs buffer layer prior to GaSe growth also potentially modifies the interface. In this case, both Ga and As atoms are present at the surface since the buffer layer is deposited at ~580°C and then lowered to the GaSe growth temperature while under As flux. The RHEED pattern of the buffer layer (see Figure 3g) exhibits much longer, well-defined streaks, indicating an atomically smooth surface[39]. Upon initiating GaSe growth by simultaneously opening the Ga and Se effusion cell shutters, the RHEED pattern immediately transitions to a tilted configuration (see Figure 3h), suggesting a change in growth orientation. The angle observed in the RHEED pattern is again $\sim 20^{\circ}$ and the GaSe layers also grow tilted in this case. Introducing a buffer layer, however, produces an atomically smooth surface for GaSe growth planes resulting in even more uniform 2D planes (Figure 4f). Comparing the AFM images in Figures. 4b, 4d, 4f clarifies the evolution of morphology. Oxide removal yields tilted ridges, adding Se during oxide removal elongates these ridges and enhances faceting, and with a buffered surface, the ridges become highly uniform and crystallized which is confirmed by XRD (Figure 5b, purple curve). Across all three substrate conditions, (a) oxide-free, (b) Se, and (c) buffered surfaces, the 2D GaSe films are tilted relative to the substrate within a growth temperature window of $300 - 500°\text{C}$. Once the native oxide is removed the film consistently adopts the tilted orientation consistent with a lattice match and bonding at the surface.

***(d) Set 4 - Native oxide layer not removed***: For the oxide prepared GaAs surface the observed RHEED pattern is a ring-shaped pattern (Figure 3j), signaling an amorphous or poorly ordered surface. Upon initiating GaSe growth on this surface, the RHEED pattern shows faint, broadened features (Figure 3k), indicative of misoriented nucleation and surface disorder[48]. This is of course a significant difference compared to sets 1-3. At the end of growth, the RHEED pattern becomes straight and streaky (Figure 3l), indicating that the film growth is potentially along the c-axis[39], in contrast to the oxide-free surface where tilted streaks are observed. The AFM topography (Figure 4h) reveals triangular, layer-like morphologies consistent with 2D spiral growth[49] attributed to the oxide chemistry in combination with roughness of the surface which promotes heterogeneous nucleation[36] and step-spiral propagation. XRD 2θ-ω scans with the diffraction vector normal to the (211) surface reveal GaSe Bragg reflections, confirming that the film adopts a (0001) orientation on the oxide-covered GaAs(211) substrate, see Figure 5c. Likewise, we also investigate the Raman spectrum to further examine the growth direction and crystal quality. As shown in Figure 5e (red curve), the film with its c-axis perpendicular to the substrate exhibits the same characteristic GaSe Raman modes observed in the tilted film of $A_{1g}^{1}{\sim}133.2\ cm^{-1}$, , $E_{2g}^{2}{\sim}207.0\ cm^{-1}$, $A_{2}^{''}{\sim}247.0\ cm^{-1}$, $E_{2g}^{1}{\sim}252.0\ cm^{-1}$, and $A_{1g}^{2}{\sim}307.0\ cm^{-1}$ [20, 28, 43, 44]. In addition, we investigated TEM to probe crystal quality and tilted growth. TEM images (see Figure 7) also directly reveal that the Set 4 of GaSe films are tilt-free. It also indicates that as the distance for growth from the interface increases, the crystal structure becomes more uniform, potentially indicating a gradual transition toward a single dominant polytype (see Figure 7b). Specifically, the γ′-GaSe polymorph[13] is locally observed using set 4 prepared substrates, see Figure 7b-c. The presence of oxide on the substrate surface may reduce interfacial selectivity and promote structural heterogeneity. Consistent with this general picture, Liu et al. showed that oxygen exposure in $MoS_2$ reduced stacking selectivity and led to mixed 3R/2H growth[50].

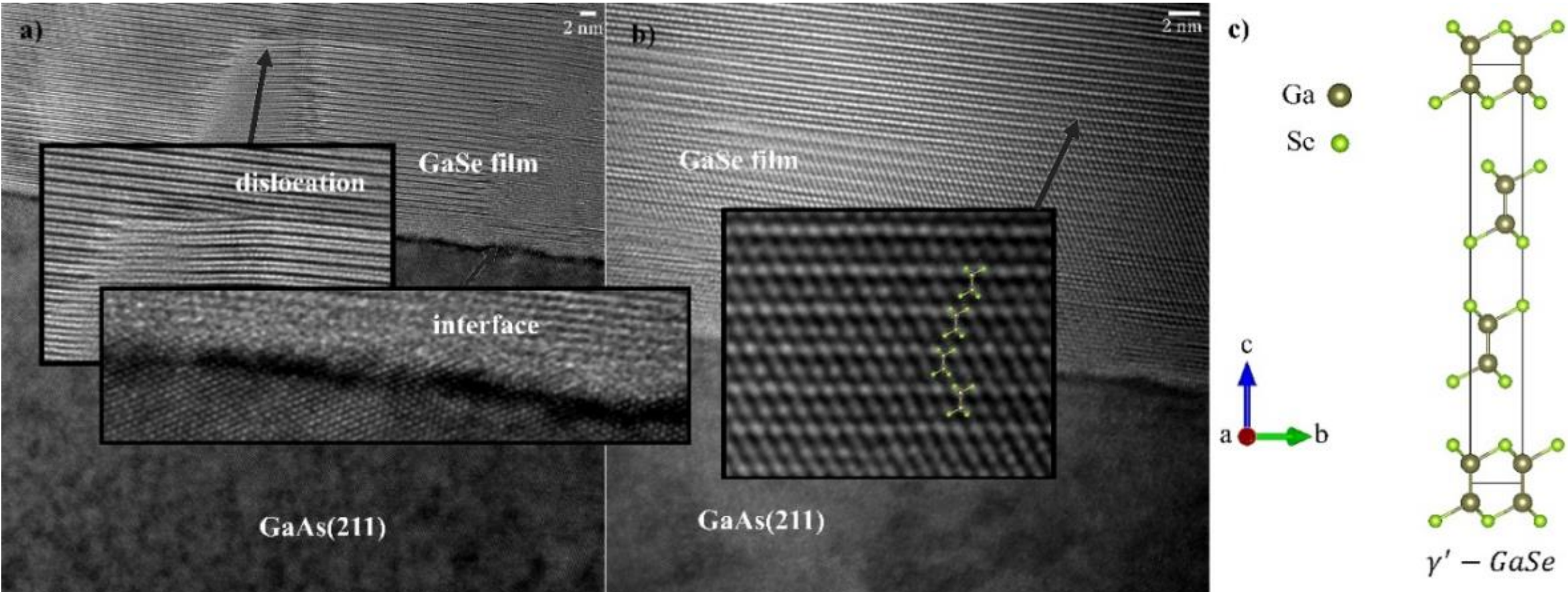

**Figure 7.** Cross-sectional TEM images of GaSe grown on oxide-covered GaAs(211). (a) TEM image showing a dislocation originating at the oxide-covered interface. The highlighted rectangular region indicates interfacial nonuniformity, where bonding is not spatially uniform; (b) High-resolution TEM image revealing the $\gamma' - GaSe$ polymorph, with the inset highlighting the atomic stacking sequence; (c) Crystal structure schematic of the $\gamma' - GaSe$ polymorph for reference.

PL measurements were also taken to evaluate the optical quality of the GaSe films grown on GaAs(211)B under various surface preparation conditions. As shown in Figure 8, all spectra are dominated by the GaAs bandgap emission near 800 nm and a weak GaSe-related feature appears in the 670-750 nm range. This GaSe-related emission is at first absent for the sample sets with tilted planes due to the inefficient optical coupling to the GaSe layers and reduced extraction of the emitted photoluminescence along the excitation direction[51]. As a result, the emission from tilted GaSe in Figure 8 is detectable only when the sample is physically tilted by approximately $\sim 20^{\circ}$ for excitation and collection normal to the planes.

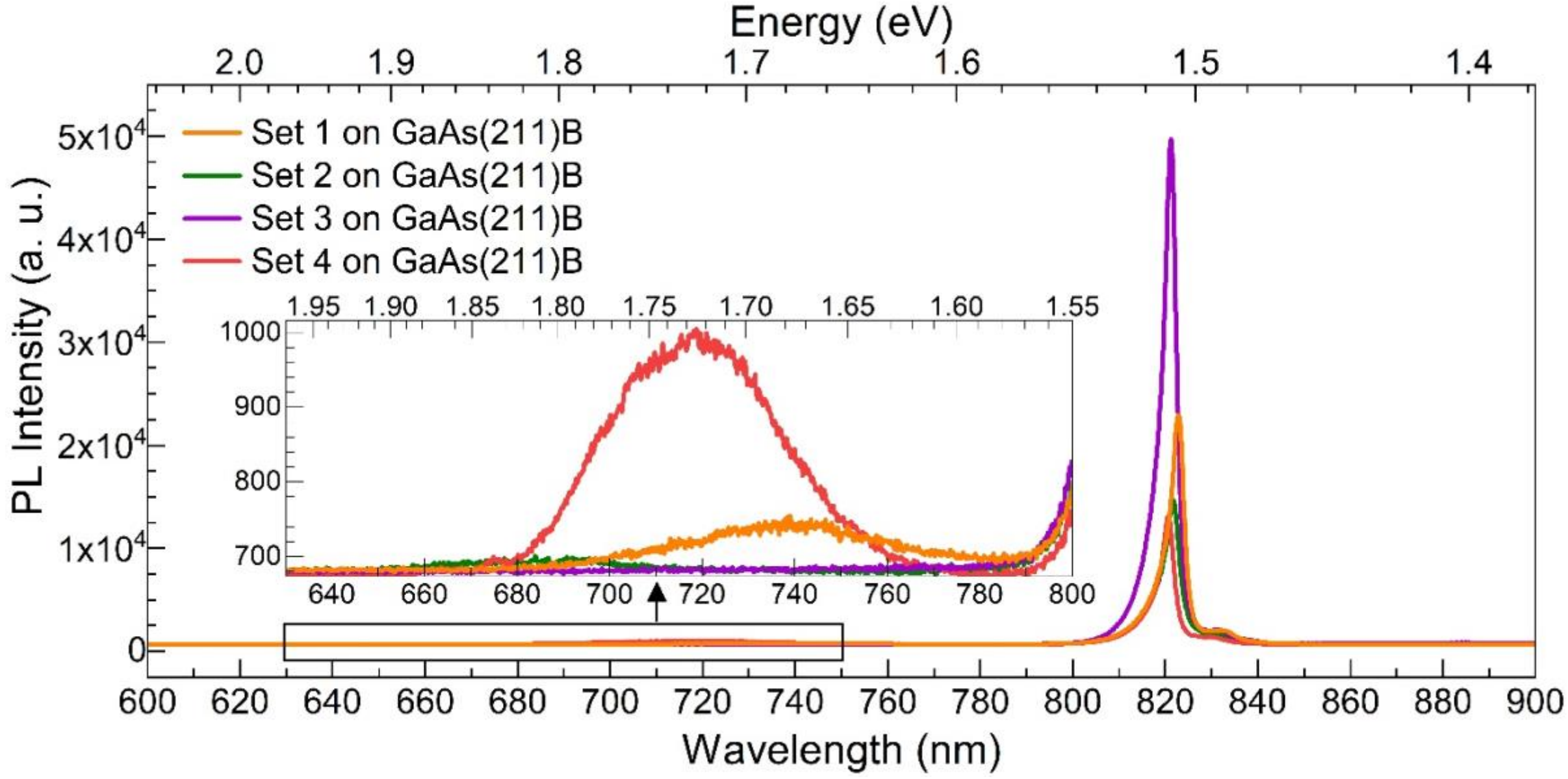

**Figure 8.** PL intensity of GaSe on oxide-free GaAs(211) (orange curve), on selenized oxide-free GaAs(211) (curve), on oxide-free buffered GaAs(211) (purple curve), and on oxide-covered GaAs(211)

In summary, our investigation of GaSe film growth by MBE on four differently prepared GaAs surfaces indicates that it is possible to grow 2D GaSe on a 3D GaAs(211) surface. However, two main trends are observed: (1) on oxide removed surfaces, due to systematic bonding, the morphology can be characterized by the growth of angled 2D GaSe planes that become more uniform over atomically smooth regions; and (2) on oxide-covered surfaces since the systematic bonding in general does not occur then the morphology can be characterized as c-axis 2D spiral-like plates spatially influenced by the variation in the substrate surface height. We speculate that if variations in the height of the substrate surface are not matched to the 2D c-axis lattice constant, nucleation at the interface can produce growth at different locations that are out of phase with one another leading to 2D spiral-like plates.

### 3.2. Growth of GaSe on GaAs(001) with different surface conditions

Based on the results for GaSe on GaAs(211), we also examined the growth of GaSe films on GaAs(001) which has also been investigated by several researchers with contrasting results.

The GaAs(001) surface is a zincblende cut perpendicular to [001]. Its tetrahedral bonds point toward the {111} family (Figure 9a), so the outermost atoms include 〈111〉 oriented dangling bonds that define chemically active sites for nucleation (Figure 9b-d).

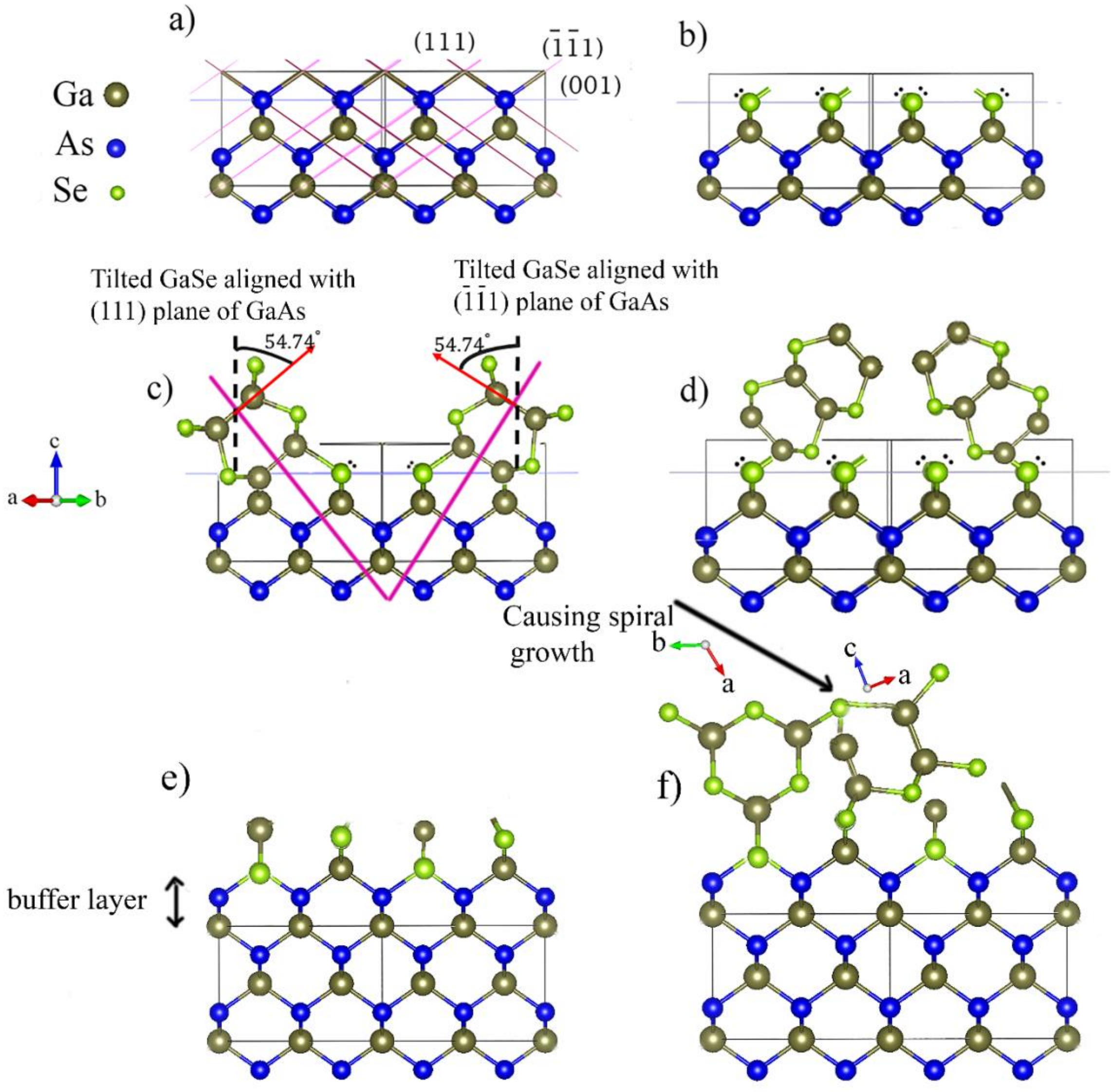

**Figure 9.** (a) Dangling bonds on the GaAs(001) As-terminated prepared surface; (b) Se prepared surface with Se atoms replacing As atoms on the surface resulting in a Se-terminated GaAs(001) surface; (c) Possible tilted growth of GaSe on the Ga-terminated surface of GaAs(001); (d) Possible tilted growth of GaSe on Se treated or Se-terminated GaAs(001) surface in (b); (e) start of growth on buffered As-terminated surface of GaAs(001) (f) growth mechanism on buffered GaAs(001) that result in spiral growth

Below, we discuss the growth of GaSe on four sets of prepared epi-ready GaAs(001) substrates*:*

***Set 1 - Oxide removed***: Before oxidation, the GaAs(001)B substrate is As-terminated, and oxidation occurs primarily at As sites, forming insulating As oxides[37]. After thermal oxide removal at $\sim 580°C$, the substrate was cooled to the growth temperature of $400°C$, resulting in a rough, Ga-terminated GaAs(001) surface, see the resulting RHEED image in Figures 10a and the morphology of the surface in Figure 11a. At the onset of GaSe deposition, Se fills the former As sites, and tilted RHEED streaks appear (Figure 10b). These streaks intensify during growth and then change to straight streaks, as shown in the final RHEED patterns in Figures 10b,c and Figure S10. This suggests that competition between growth in both directions ultimately leads to coalescence into growth normal to the surface as shown in AFM image in Figure 11b. On the GaAs(001) surface, each surface Ga atom presents two dangling bonds oriented towards two of the four $\langle 111 \rangle$ directions. Please refer to Figure S11 for the geometry of Ga-terminated GaAs(001). This geometric arrangement naturally selects two of the four symmetry-equivalent variants, so GaSe nucleates on the surface in two tilted orientations which of course can cause competition in growth. This interpretation is supported by XRD data showing that GaSe reflections are observed when the diffraction vector is normal to the surface and also when the diffraction vector is normal to (111) plane in GaAs(001) substrate (the sample is tilted by $\sim 54^{\circ}$ in either direction in the (110) plane (see Figure 12a(1) and (2) ) and Figure S12), This motivated us to investigate samples with much lower depositions thickness than those shown in the Figure 11b, for which we more clearly observe tilted growth at $\sim 54^{\circ}$ in both directions from the normal (inset of Figure 11b, see the enlarged photo of tilted growth in Figure S13), indicating that the GaSe (0001) axis is inclined relative to the surface normal of the (001)-oriented GaAs substrate. In this case, the 2D GaSe (0001) planes are aligned with two of the substrate's $\langle 111 \rangle$ directions. The result indicates that depending on growth thickness one can observe either tilted growth or growth perpendicular to the surface.

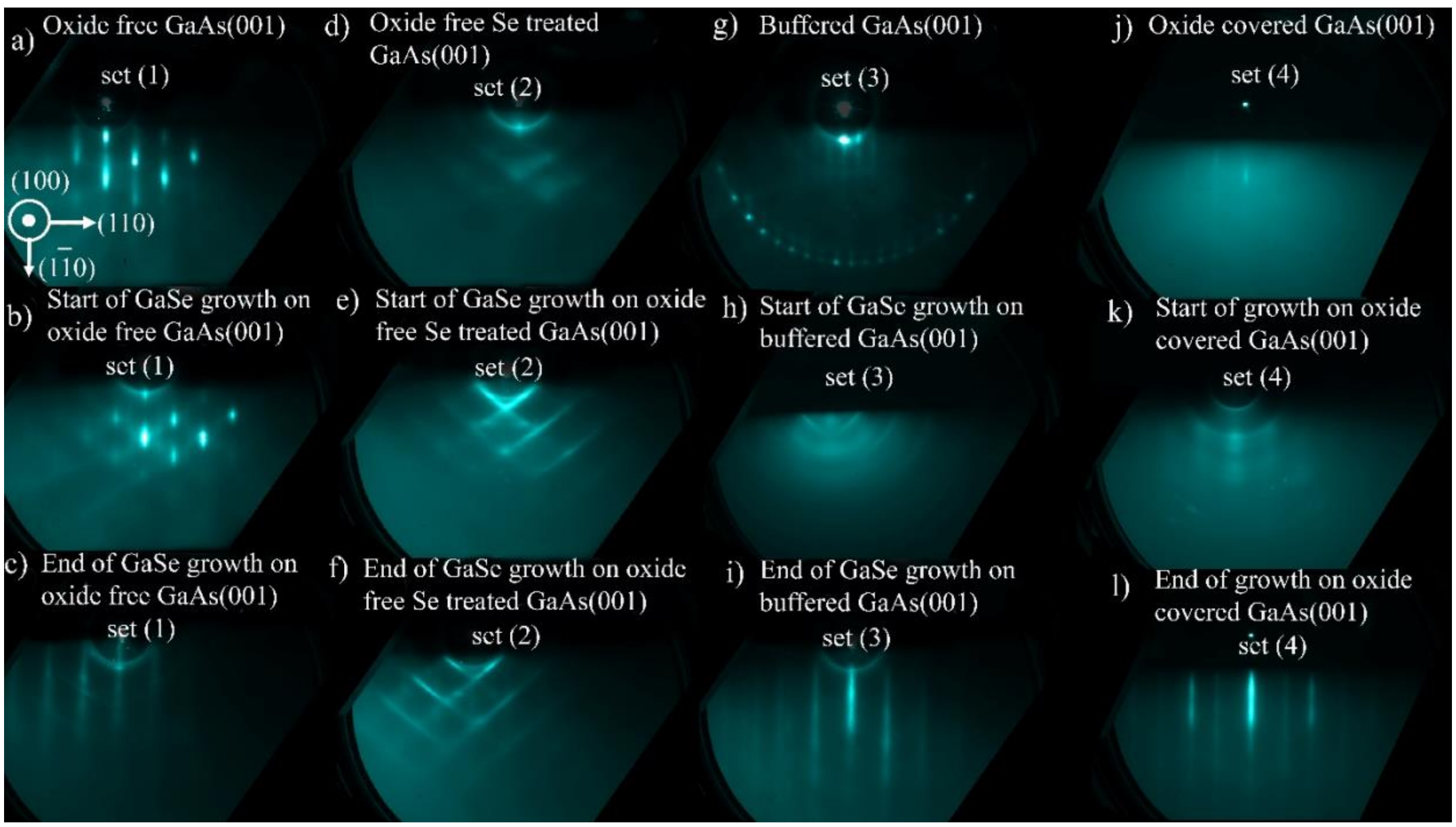

**Figure 10.** RHEED patterns of GaAs(001) substrates and GaSe films under various surface preparation conditions: (a) Oxide-free GaAs(001) substrate; (b) Beginning of GaSe growth on oxide-free GaAs(001) substrate; (c) End of growth of GaSe on oxide-free GaAs(001) substrate; (d)Se treated oxide-free GaAs(001); (e) Beginning of GaSe growth on Se treated oxide-free GaAs(001) substrate; (f) End of GaSe growth on Se treated oxide-free GaAs(001) substrate; (g) Buffered GaAs(001) substrate; (h) Beginning of GaSe growth on buffered GaAs(001) substrate; (i) End of growth of GaSe on buffered GaAs(001) substrate; (j) Oxide-covered GaAs(001) substrate; (k) Beginning of GaSe growth on oxide-covered GaAs(001) substrate; (l) End of growth of GaSe on oxide-covered GaAs(001) substrate.

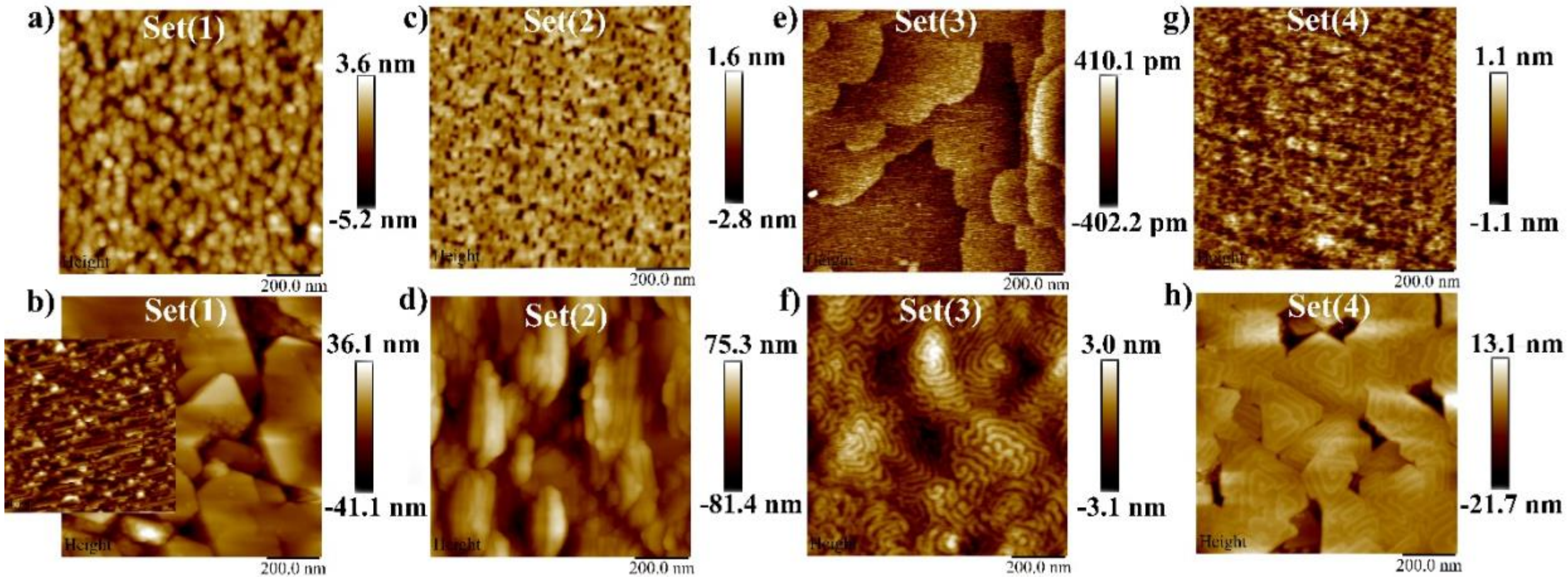

**Figure 11.** Atomic force microscopy of GaAs(001) substrates and GaSe films under various surface preparation conditions: (a) Oxide-free GaAs(001) substrate; (b) GaSe film on oxide-free GaAs(001) substrate; (c) Se treated Oxide-free GaAs(001) substrate; (d) GaSe film growth on Se treated oxide-free GaAs(001) substrate; (e) Buffered GaAs(001) substrate; (f) GaSe film growth on buffered GaAs(001) substrate; (g) Oxide-covered GaAs(001) substrate; (h) GaSe film growth on oxide-covered GaAs(001) substrate. All images are 1 μm × 1 μm in size, and the scale bar in each panel represents 200 nm.

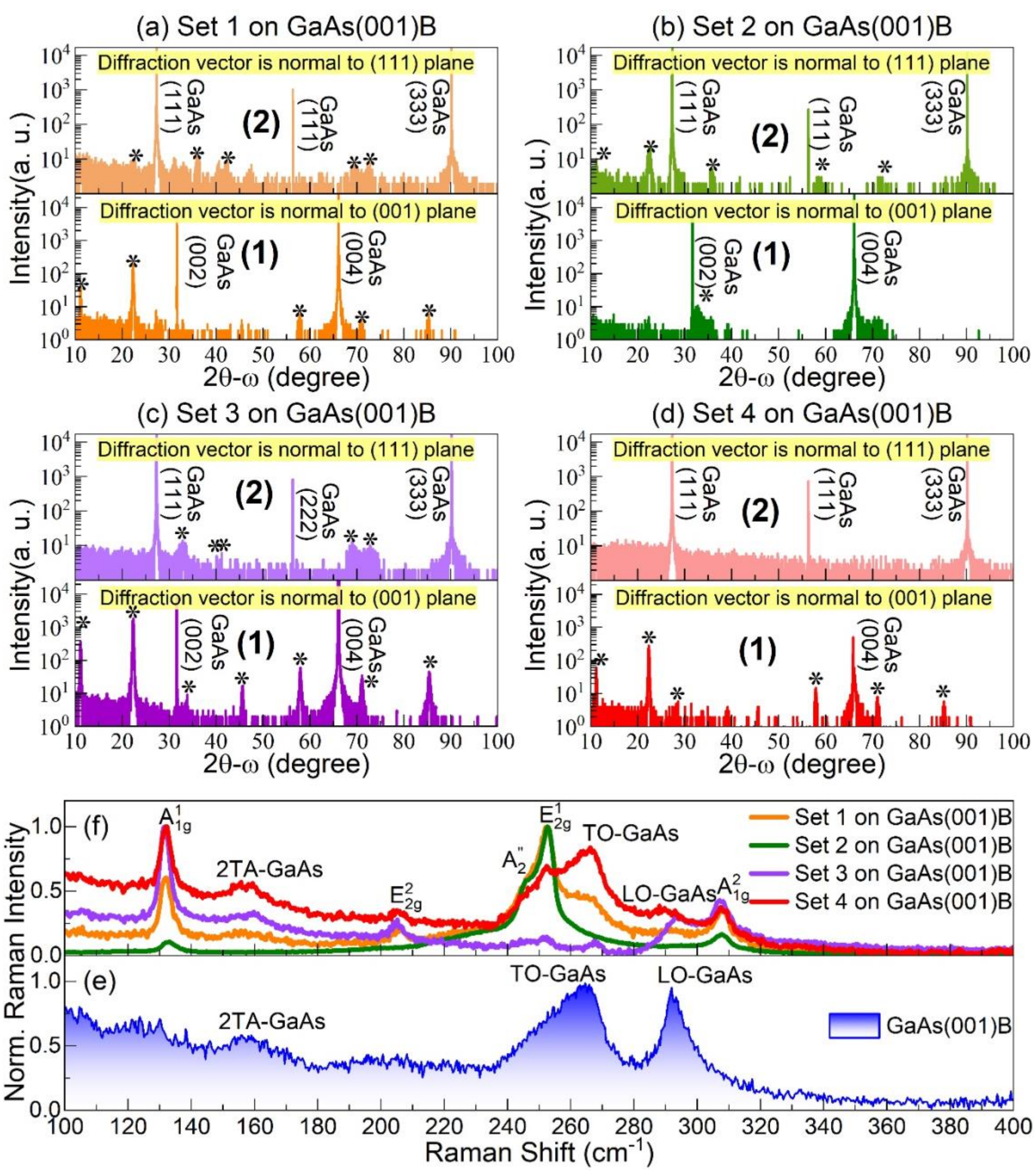

**Figure 12.** XRD patterns of GaSe/GaAs(001) films with varying surface preparations. (a) Oxide-free; (b) selenized; (c) buffered; (d) oxide-covered. For each condition, scans were collected along (001) plane of GaAs(001), labeled as (1), and GaAs(111), labeled as (2). Asterisks indicate GaSe peaks. (e) Raman spectrum of the GaAs(001)B substrate. (f) Raman spectra of GaSe films grown on oxide-free (set-1, orange curve), selenized oxide-free (set-2, green curve), buffered (set-3, purple curve), and oxide-covered (set-4, red curve), as well as GaAs(001)B substrate (blue curve).

***Set 2 - Oxide removed under Se flux**:* We examined GaSe films grown on GaAs(001) with surface selenization. The RHEED patterns recorded before, during, and after growth show intense tilted streaks for the selenized surface (Figures. 10d–f) (Supporting Information, Figure S14 and Figure S15). AFM also reveals that alignment along two specific ⟨111⟩ directions of the GaAs(001) substrate (Figure 11d). We attribute this to differences in surface termination. A Ga-terminated GaAs(001) surface presents two dangling bonds per atom, enabling both Ga and Se to bond, with Ga-Se bonding energetically favored (due to electronegativity difference), thereby allowing multiple tilted variants at different locations. In contrast, a Se adlayer provides only one dangling bond per Se atom, restricting nucleation to Ga-Se bonds and reducing the number of accessible variants and competition (due to Se bonding in GaSe which only bonds with Ga). Consequently, the Se-terminated surface favors a selective tilted growth mode consistent with the RHEED observations and XRD (Figs. 10c, 10f, 12b(1), 12b(2) and Supporting Information Figure S14 and Figure S15).

***Set 3 - Oxide removed and GaAs buffer added**:* In this case, the oxide is again removed at approximately $\sim 580$°C in the presence of As, producing an As-terminated GaAs(001) surface. The sample is then cooled to the GaSe growth temperature under As flux, resulting in an As-terminated surface. The As atoms on the surface can bond to both Ga and Se with comparable probability because the electronegativity differences for As–Ga and As–Se are similar, giving rise to a chemically heterogeneous interface, see Figure 9e and Figure 9f. The RHEED pattern along the [110] azimuth further shows that growth begins with tilted GaSe layers at $\sim 54^{\circ}$, followed by a transition to c-axis-oriented GaSe aligned perpendicular to the (001) plane (Figure 10h-Figure 10i). In particular, when As bonds with Ga, the associated dangling-bond configuration is effectively rotated by $90^{\circ}$(bonding situation in GaAs), whereas bonding involving Se does not impose the same geometric relationship. This interfacial mismatch can promote growth along multiple in-plane directions, consistent with the spiral-like morphology observed by AFM (Figure 11f) and the presence of multiple growth orientations indicated by the supplementary XRD results (See Figure S14). This interpretation is further supported by the XRD measurements in Figure 12c(1) and Figure 12c(2), where GaSe Bragg reflections are observed when the diffraction vector is normal to the (001) plane as well as to the four {111} planes.

The quality of the films is again confirmed by the characteristic Raman-active modes of GaSe ($A_{1g}^{1}$~$133.2\ cm^{-1}$, $E_{2g}^{2}$~$207.0\ cm^{-1}$, $A_{2}^{''}$~$246.0\ cm^{-1}$, $E_{2g}^{1}$~$252.0\ cm^{-1}$, $A_{1g}^{2}$~$307.0\ cm^{-1}$), as expected.

***Set 4 – Native oxide layer not removed***: For GaSe films grown by MBE on a (001)-oriented GaAs substrate with its native oxide layer, our observation is similar to those for growth on a GaAs(211) substrate with its native oxide layer. The initial RHEED pattern is ring-shaped (Figure 10j), which evolves into straight streaks by the end of growth (Figure 10l). This observation, together with the AFM topography images in Figure 11h, reveals the triangular helicoidal features that were also observed previously. Meanwhile, the XRD pattern in Figure 12d, confirms that GaSe films adopt a (0001) orientation without tilt growth. Raman spectroscopy in Figure 12f shows the characteristic modes of GaSe. As shown in Figure 13, we also investigated the PL from the four sample sets. Similarly to the PL seen in Figure 8, the observed emissions covered different spectral regions for substrates that were prepared under different surface conditions. Tilted growth samples tended to have a higher energy component to the spectrum, which we speculate can be due to the thickness of the 2D layer[52].

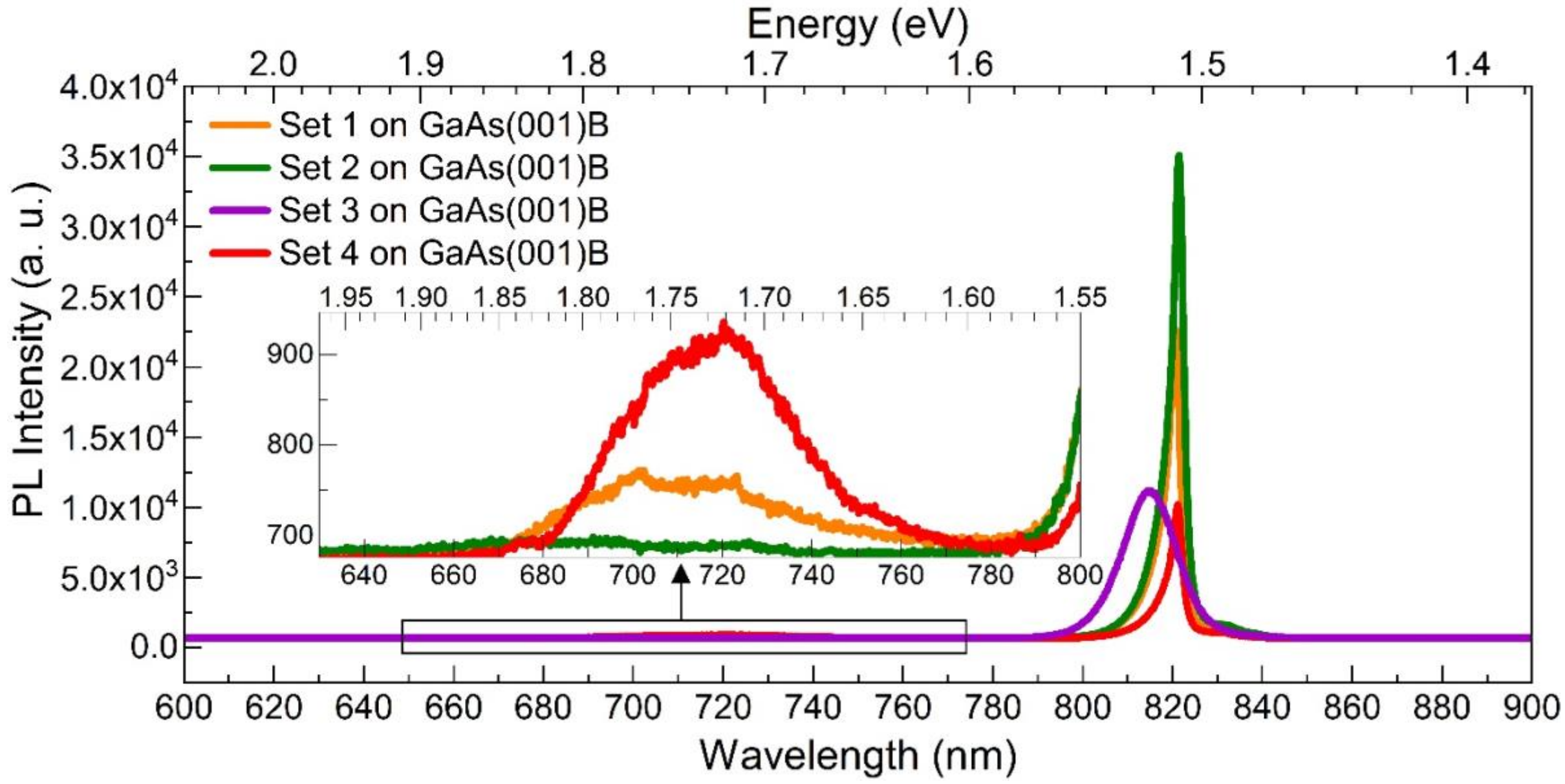

**Figure 13.** PL intensity of GaSe on oxide-free GaAs(001), selenized GaAs(001), buffered GaAs(001), and oxide-covered GaAs(001).

In summary, the growth of GaSe on GaAs(001) is more complicated than growth on GaAs(211) due to competition for nucleation in two and four different directions. However, the outcomes show two main trends: (1) if systematic bonding occurs, the morphology is characterized by the growth of angled 2D GaSe planes that can be more uniform over atomically smooth regions. However, there can be competition between growth at different angles resulting in a compromised growth along the c-direction and in set 3 growth characterized by 2D spiral plates. (2) If systematic bonding does not occur then the morphology will be characterized by 2D spiral-like plates. In all cases, the morphology of growth is highly dependent on the atomic smoothness and step nature of the surface, as might be expected for 2D layer growth. This can explain the differences in previous reports for growth on GaAs(001). Although no details are given here, we also probed growth on the high index GaAs surfaces of GaAs(311) and (411) and found similar and consistent results briefly summarized in Table 1 and Table S1 in Supporting Information.

**Table 1.** GaSe growth modes on GaAs(211) and GaAs(001) under different interfacial conditions

| *Substrates* | *Oxide removed* | *Se treated* | *Buffered* | *Oxide-covered* |
|---|---|---|---|---|
| GaAs(211) | Growth angle (19.46°) | Growth angle (19.46°) | Growth angle (19.46°) | 2D spiral growth |
| GaAs(001) | Growth angle (54.73°) | Growth angle (54.73°) | 2D spiral growth | 2D spiral growth |

## 4. Conclusion

Our results reveal how substrate preparation and subsequent bonding of the GaSe film with the substrate explain the different outcomes reported in previous studies and add insight on the mechanism for growth of a 2D material on a 3D substrate. By systematically comparing GaAs(211)B and GaAs(001)B substrates under controlled surface chemistries, including native oxide, oxide removal, Se pre-exposure, and GaAs buffer layers, we directly identify the interfacial mechanisms that select growth mode, crystallographic orientation, and evolution in 2D/3D heterostructures.

More broadly, this work provides the mechanistic framework describing how interfacial bonding and surface symmetry control layered 2D/3D heteroepitaxy. Our observations are consistent with previous investigations of the growth of 2D materials on 3D substrates[53, 54]. Our finding emphasizes more generally that if covalent bonding occurs at periodic local positions along the 2D-3D interface, growth of the 2D material will follow at an angle to the substrate set by the bonding angle[54, 55]. On the other hand, if local bonding is for some reason prevented from occurring, for example, by oxidation, hydrogenation, or some other type of surface passivation, then van der Waals growth will follow normal to the surface. What has not always been clear in previous works is if the substrate surface was passivated, or if the atomic spacing of the substrate and 2D material allows local bonding. This has resulted in conflicting reports in the literature [26-28, 56]. The results presented here clarify the origin of these previously conflicting observations.

## Supporting Information

Film thickness information, angle analysis, supplementary RHEED patterns, AFM line profiles and morphology, XPS spectra and elemental ratios, polarization-dependent Raman measurements, additional XRD scans, temperature-dependent growth results, and comparison of GaSe growth orientations on GaAs substrates with different crystallographic orientations.

## Author Information

Corresponding Author

Aida Sheibani: Smart Ferroic Materials center, Physics Department, University of Arkansas, Fayetteville, AR 72701, USA. ORCID:0009-0008-2173-4683
Email: asheiban@uark.edu

**Authors:**

Mohammad Zamani-Alavijeh: Institute for Nanoscience and Engineering, University of Arkansas, Fayetteville, AR 72701, USA.ORCID:0000-0002-3742-2499

Charles Paillard: Smart Ferroic Materials center, Physics Department, University of Arkansas, Fayetteville, AR 72701, USA. Université Paris-Saclay, CentraleSupélec, CNRS, Laboratoire SPMS, 91190, Gif-sur-Yvette, France.ORCID: 0000-0002-3337-2295.

Kanagaraj Moorthi: Institute for Nanoscience and Engineering, University of Arkansas, Fayetteville, AR 72701, USA. ORCID: 0000-0001-7339-714X.

Fernando Maia de Oliveira: Institute for Nanoscience and Engineering, University of Arkansas, Fayetteville, AR 72701, USA. ORCID: 0000-0003-4272-5416

Serhii Kryvyi: Institute for Nanoscience and Engineering, University of Arkansas, Fayetteville, AR 72701, USA. ORCID: 0000-0003-3202-3154.

Mourad Benamara: Institute for Nanoscience and Engineering, University of Arkansas, Fayetteville, AR 72701, USA. ORCID: 0000-0003-1408-8202.

Hryhorii Stanchu: Institute for Nanoscience and Engineering, University of Arkansas, Fayetteville, AR 72701, USA. ORCID: 0000-0002-2987-1251.

Calbi Gunder: Air Force Research Laboratory, 2241 Avionics Circle, Wright-Patterson AFB, OH 45433, USA. ORCID: 0000-0001-6112-3061.

Hugh Churchill: Smart Ferroic Materials center, Physics Department, University of Arkansas, Fayetteville, AR 72701, USA. Institute for Nanoscience and Engineering, University of Arkansas, Fayetteville, AR 72701, USA. ORCID: 0000-0002-8287-1373.

Yuriy I. Mazur: Institute for Nanoscience and Engineering, University of Arkansas, Fayetteville, AR 72701, USA. ORCID: 0000-0002-0884-6049.

Gregory Salamo: Smart Ferroic Materials center, Physics Department, University of Arkansas, Fayetteville, AR 72701, USA. Institute for Nanoscience and Engineering, University of Arkansas, Fayetteville, AR 72701, USA. ORCID: 0000-0002-5962-784X.

Author Contributions

Aida Sheibani: Performed GaSe molecular beam epitaxy growth, carried out AFM measurements, analyzed the experimental data, prepared figures, and wrote the initial manuscript draft.

Mohammad Zamani-Alavijeh: Contributed to GaAs buffer growth, supported MBE laboratory operation, and performed XPS measurements.

Charles Paillard: Contributed to scientific advising, interpretation of the growth mechanism, and manuscript writing and revision.

Kanagaraj Moorthi: Contributed to GaSe growth calibration and experimental support.

Fernando Maia de Oliveira: Performed Raman and photoluminescence measurements and contributed to optical data analysis and manuscript revision.

Serhii Kryvyi: Performed X-ray diffraction measurements and contributed to structural characterization.

Mourad Benamara: Contributed to SEM and TEM sample preparation, measurements, and structural analysis.

Hryhorii Stanchu: Contributed to SEM and TEM sample preparation, measurements, and structural analysis.

Calbi Gunder: Contributed to GaAs buffer calibration, experimental discussion, and manuscript revision.
Hugh Churchill: Contributed to scientific advising, interpretation of two-dimensional semiconductor integration, and manuscript revision.

Yuriy I. Mazur: Contributed to the design and implementation of Raman and photoluminescence measurements and optical characterization.

Gregory Salamo: Supervised the project, guided the experimental design and interpretation, coordinated the research effort, and revised the manuscript.

## ACKNOWLEDGEMENTS

The authors acknowledge partial support from $\mu$-ATOMS, an Energy Frontier Research Center funded by the U.S. Department of Energy (DOE), Office of Science, Basic Energy Sciences (BES), under award DE-SC0023412, a grant from the Air Force Office of Scientific Research through award no. FA9550-24-1-0263, and the National Science Foundation under Grant No. DGE-2244274. We also acknowledge the MonArk NSF Quantum Foundry supported by the National Science Foundation Q-AMASE-i program under NSF award No. DMR-1906383.

## CONFLICT OF INTEREST

The authors declare that they have no conflict of interest.

TOC

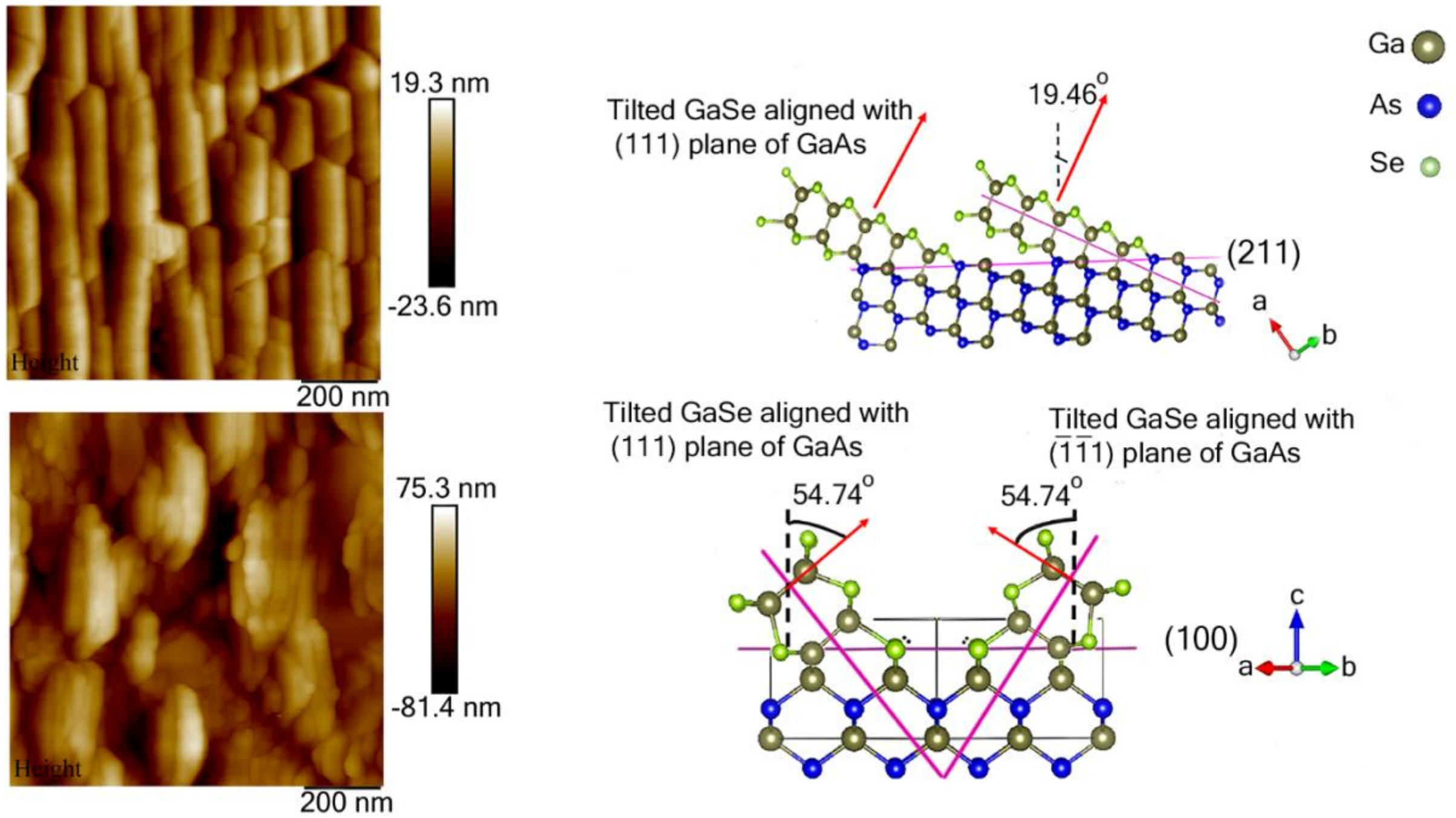